\documentclass[mypaper,8pt,twoside]{CoAst}
\usepackage{epsf,graphicx,fancyhdr}
\renewcommand{\chaptermark}[1]{\markboth{#1}{}}

\newcommand{\kopf}{\small\itshape Comm. in Asteroseismology \\ Contribution to the Proceedings of the Wroclaw HELAS Workshop, 2008}

\newcommand{\Authors}[1]{\begin{center}\normalsize\bf\sf #1 \end{center}}

\renewcommand{\author}[1]{\begin{center}\normalsize\bf\sf #1 \end{center}}
\newcommand{\Address}[1]{\begin{center}\small\sf #1 \end{center}}

\newcommand{\Session}[1]{{\vspace{3mm}\small \noindent  \hspace*{3mm} Session: } #1 \normalsize}

\newcommand{\Objects}[1]{{\vspace{0mm}\small \noindent  \hspace*{3mm} Individual Objects: } \small #1 \normalsize}

	\newcommand{\oneA}{\small DATA - ground-based photometry \newline}

\renewenvironment{abstract}{\section*{Abstract}\normalsize\sf}{}
\newcommand{\References}[1]{\begin{flushleft}{\large References\\}\vspace*{2mm}\small #1 \end{flushleft}}

\newcommand{\chapterCoAst}[2]{\chapter[\sf\normalsize #1\\ \footnotesize \hspace*{5mm}by #2 \sf\normalsize][]{#1\\}\rhead[\fancyplain{}{\sf\footnotesize \center{#1}}]{\fancyplain{}{\sffamily\thepage}}\lhead[\fancyplain{\kopf}{\sffamily\thepage}]{\fancyplain{\kopf}{\sf\footnotesize \center{#2}}}}

\newcommand{\figureDSSN}[5]{\begin{figure}[#4]
\centering
\includegraphics*[#5]{#1}
\caption{#2}
\label{#3}
\end{figure}}

\renewcommand{\chaptername}{}

\newcommand{\acknowledgments}[1]{\vspace*{5mm}\noindent  \textbf{Acknowledgments.} #1}

\def\rfr{\smallskip\par\noindent
        \hangindent=7truemm
        \hangafter=1}

\sf
\begin{document}

\chapterCoAst{Regularities in the frequency spacings of\\Delta Scuti stars and the s-f Diagram}
{M. Breger, P. Lenz, A. A. Pamyatnykh}

\Authors{M.~Breger$^1$, P.~Lenz$^1$, A. A. Pamyatnykh$^{1,2}$} 
\Address{$^1$ Astronomisches Institut der Universit\"at Wien, T\"urkenschanzstr. 17, A--1180 Wien, Austria\\
$^2$ Copernicus Astronomical Center, Bartycka 18, 00-716 Warsaw, Poland}

\noindent
\begin{abstract}

Statistical analyses of several $\delta$ Scuti stars (FG Vir, 44 Tau, BL Cam and others) show that the photometrically observed frequencies cluster around the frequencies of the radial modes over many radial orders. 

The observed regularities can be partly explained by modes trapped in the stellar envelope. This mode selection mechanism was already proposed by Dziembowski \& Kr\'{o}likowska (1990) and was shown to be efficient for $\ell=1$ modes. New pulsation model calculations confirm the observed regularities.

We present the s-f diagram, which compares the average separation of the radial frequencies ($s$) with the frequency of the lowest unstable radial mode ($f$). The diagram provides an estimate for the $\log g$ value of the observed star, if we assume that the centers of the observed frequency clusters correspond to the radial mode frequencies. This assumption is confirmed by examples of well-studied $\delta$~Scuti variables in which radial modes were definitely identified.

\end{abstract}

\Session{ \oneA}
\Objects{FG\,Vir, 44\,Tau, BL\,Cam} 

\section{Introduction}
Recent observational campaigns carried out with earth-based telescopes or space missions concentrating on selected stars have revealed a rich spectrum of radial and nonradial modes covering a wide range in frequencies. This range in frequencies varies from star to star and depends on a variety of factors, not all of which are understood. The question of which modes are selected by the star is not solved at the present time.

The question arises of whether the mixture of the excited radial and nonradial modes shows frequencies which are essentially randomly distributed over the range of unstable frequencies or whether they tend to form groups. An example of the latter is the regular spacing found in high-order nonradial pulsation (the asymptotic case), as detected in the Sun and white dwarfs.
The $\delta$ Scuti stars, on the other hand, pulsate with low-order p (and g) modes. Here we examine the frequency distribution of mostly nonradial
modes in a number of well-observed $\delta$ Scuti stars in order to search for regularities.

\section{Regularities in the frequency distribution}

The detection of the complex mixture of excited nonradial and (a few) radial modes observed in $\delta$ Scuti stars requires extensive photometric studies. Sufficient information for meaningful statistical analyses concerning regularities in the frequency spacings is only available for a few stars. We find that three well-studied stars show pronounced regularities:

(i) FG Vir is probably the best $\delta$ Scuti star to examine systematics in the pulsation frequencies. More than 75 frequencies of pulsation are known (Breger et al. 2005).
The frequency resolution is excellent so that combination frequencies and harmonics can be eliminated; this leaves 68 independent frequencies covering
a wide range from 5.7 to 44~cd$^{-1}$. The regions with most frequencies are around 12, 23 and 34 cd$^{-1}$, but the data contain considerably more information than a spacing of 11 cd$^{-1}$, which corresponds to three radial orders. A statistical analysis shows a pronounced regularity with a spacing of  3.7 cd$^{-1}$, which corresponds to the average spacing of consecutive radial orders for these mainly nonradial modes.

(ii)  Extensive campaigns of 44 Tau covering five observing seasons have led to the detection of 49 frequencies (Breger \& Lenz 2008). As already found in FG Vir, the frequencies of 44 Tau also show a regularity with a preferred spacing of one radial order.

(iii)  The star BL Cam (Rodr\'iguez et al. 2007). While the two stars listed above are small-amplitude $\delta$ Scuti variables, BL Cam is a high-amplitude, extremely metal-deficient variable (also known as HADS, SX Phe-type). Rodr\'iguez et al. (2007) identified 25 frequencies, of which 22 represent independent modes. 
The study is remarkable because of the difficulty of detecting such a large number of small-amplitude nonradial modes in the presence of a dominant radial fundamental mode of high amplitude. The authors note that the frequencies of the nonradial modes cluster in groups near 25, 32, 46, 51--53, and 72--80 cd$^{-1}$. This implies separations similar to the separation of adjacent radial orders.

\section{Frequencies of nonradial modes near those of radial modes}

In the previous section, we have shown that in a number of well-studied $\delta$ Scuti stars, the frequencies of the nonradial modes are not distributed at random but show a preferred spacing corresponding to that of the radial modes. We wish to emphasize that this is only a preferred spacing and that other spacings do (and should) occur. 
The question arises of where in the frequency spectrum these concentrations of nonradial modes occur, e.g., possibly halfway between radial modes of successive radial orders, as would be expected in the asymptotic case for the $\ell$ = 1 modes.  For the three stars, at least one radial mode has been identified in each star.
We have computed the frequencies of the other radial modes from the known
properties of the stars.  The same codes as described in Lenz et al. (2008) were used. Knowing the radial frequencies, it was possible to
compute the frequency difference of each observed mode to the nearby radial modes.

\figureDSSN{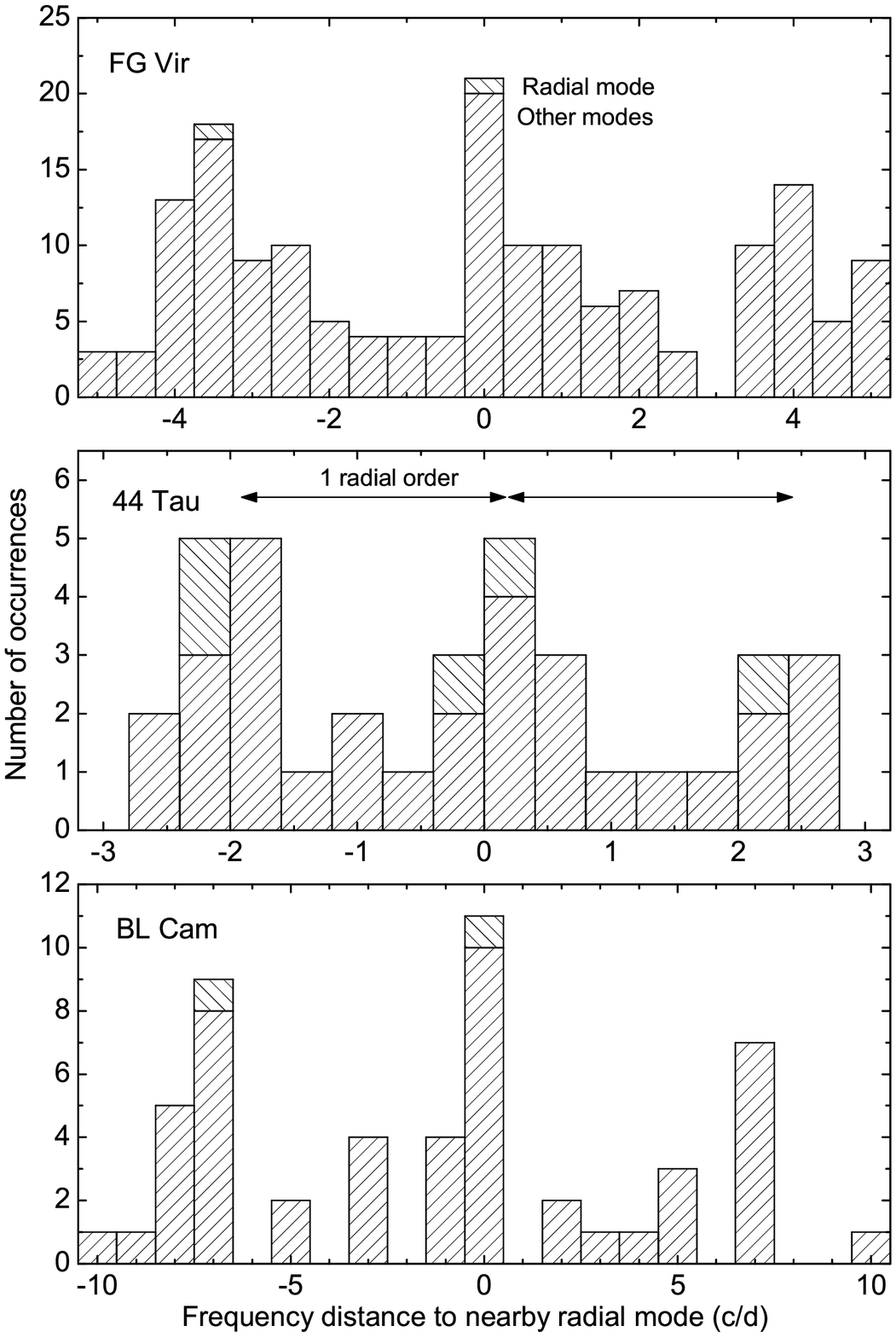}{Histogram of the frequency distances of individual modes to the frequency of the nearest radial mode. The frequencies of the radial modes were either
observed or computed from models. This diagram shows that the observed nonradial modes are not distributed at random but tend to cluster around the radial modes.}{label}{!ht}{bb=25 40 530 720,width=90mm,clip}

The histograms of the frequency differences are shown for the three stars in Fig. 1. The results are striking:
The frequencies of the photometrically detected nonradial modes are not distributed at random but tend to cluster around those of the radial modes. These nonradial modes are mostly $\ell$ = 1 and 2 modes, but in the case of FG Vir they may also be the small-amplitude $\ell$ = 3 and 4 modes.

Three additional $\delta$ Scuti stars (XX Pyx, BI CMi and $\epsilon$ Cep) with 20+ known frequencies (i.e., borderline statistics) were also examined. They are less ideal for the study because of  close-binary nature or uncertain physical parameters. The stars show the radial-spacing regularities, but less pronounced than the three stars examined in the previous section.

We conclude that in $\delta$ Scuti stars, the detected nonradial modes tend to cluster around the radial modes. This effect has previously been predicted:
Dziembowski \& Kr\'{o}likowska (1990) examined {\em mode trapping} in the stellare envelope as a mechanism for mode selection in $\delta$ Scuti stars. They show that some modes of $\ell=1$ are trapped in the envelope and, therefore, are less coupled to g modes in the deep interior. They have a higher probability to be excited to observable amplitudes than other modes. Trapped modes are nonradial counterparts of the acoustic radial modes and at low spherical harmonic degrees their frequencies are close to those of the radial modes. Mode trapping is not effective for $\ell=2$ modes. However, the observations show that even modes identified with $\ell=2$ are located close to the radial modes. This means that an additional mode selection might exist.

If the star rotates, its oscillation frequencies split into multiplets which disturb the grouping of the modes around radial frequencies. We tested this effect and found that at rotational velocities less than 100 km/s the regularities in the frequency spectra still remain easily detectable. Moreover, mostly axisymmetric modes are observed, which are only slightly influenced by rotation.

\figureDSSN{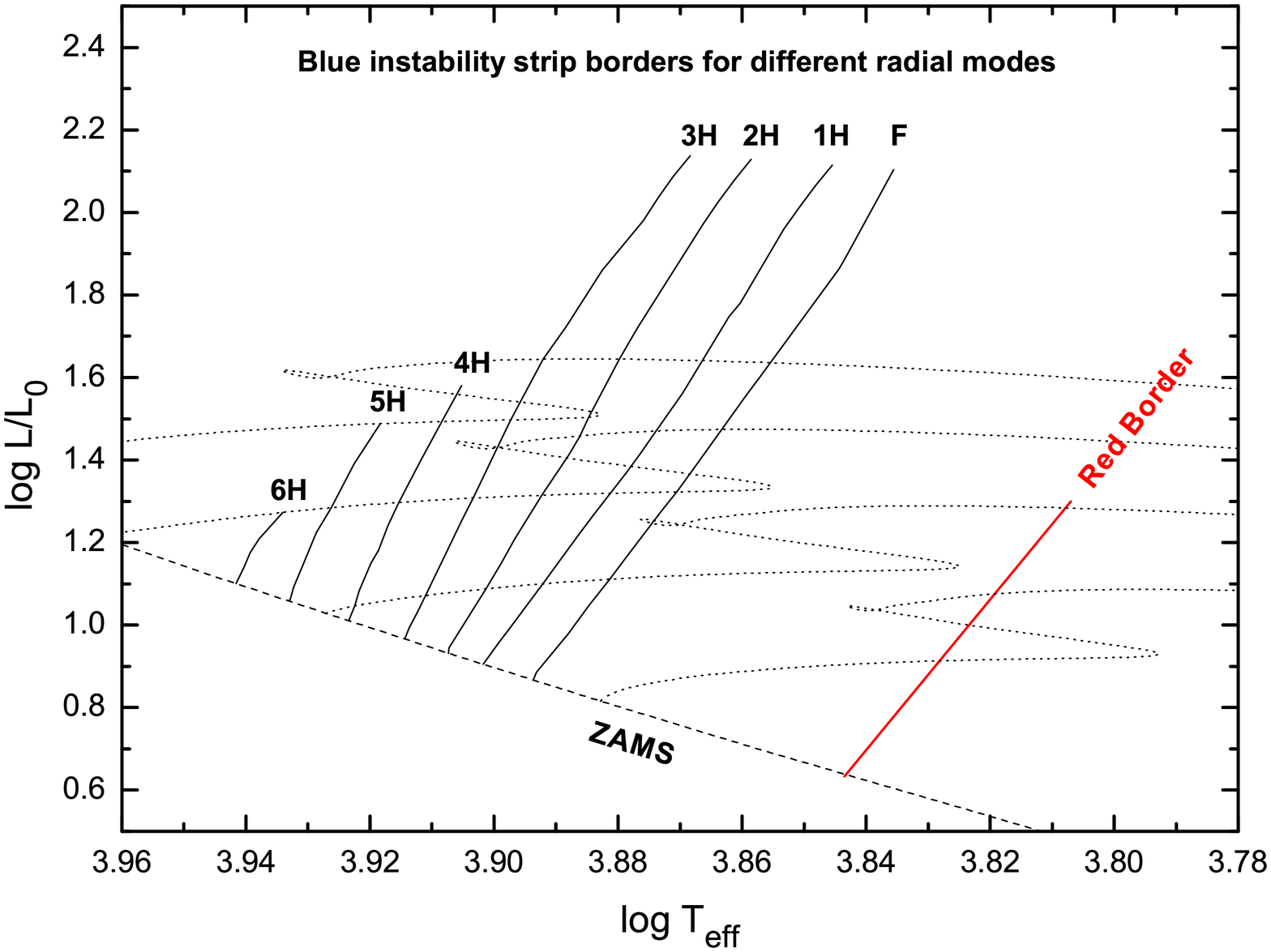}{Predicted blue instability strip borders of the radial modes. Evolutionary tracks for models with 1.6, 1.8, 2.0 and  2.2 M$_\odot$ are given by dotted lines. The red border was taken from Dupret et al. (2004).}{label}{!ht}{bb=15 15 800 620,width=100mm,clip}

\section{Application of Regularities: The s-f diagram}

We shall now assume that the centers of the observed frequency clusters correspond to the frequencies of the radial modes. Our observations of the stars examined above show that this assumption is almost satisfied. With this assumption the presence of regularities in observed frequency spectra may be used to infer fundamental parameters of stars in the $\delta$~Scuti domain in the HRD if these parameters are uncertain or unknown.

We followed the evolution of a star and computed radial frequencies of the models. The same codes as described in Lenz et al. (2008) were used. A detailed inspection of the results reveals an excellent possibility to determine the $\log g$ value of a star by means of two parameters: the average frequency separation between the radial modes, $s$, and the frequency of the lowest unstable radial mode, $f$.

\figureDSSN{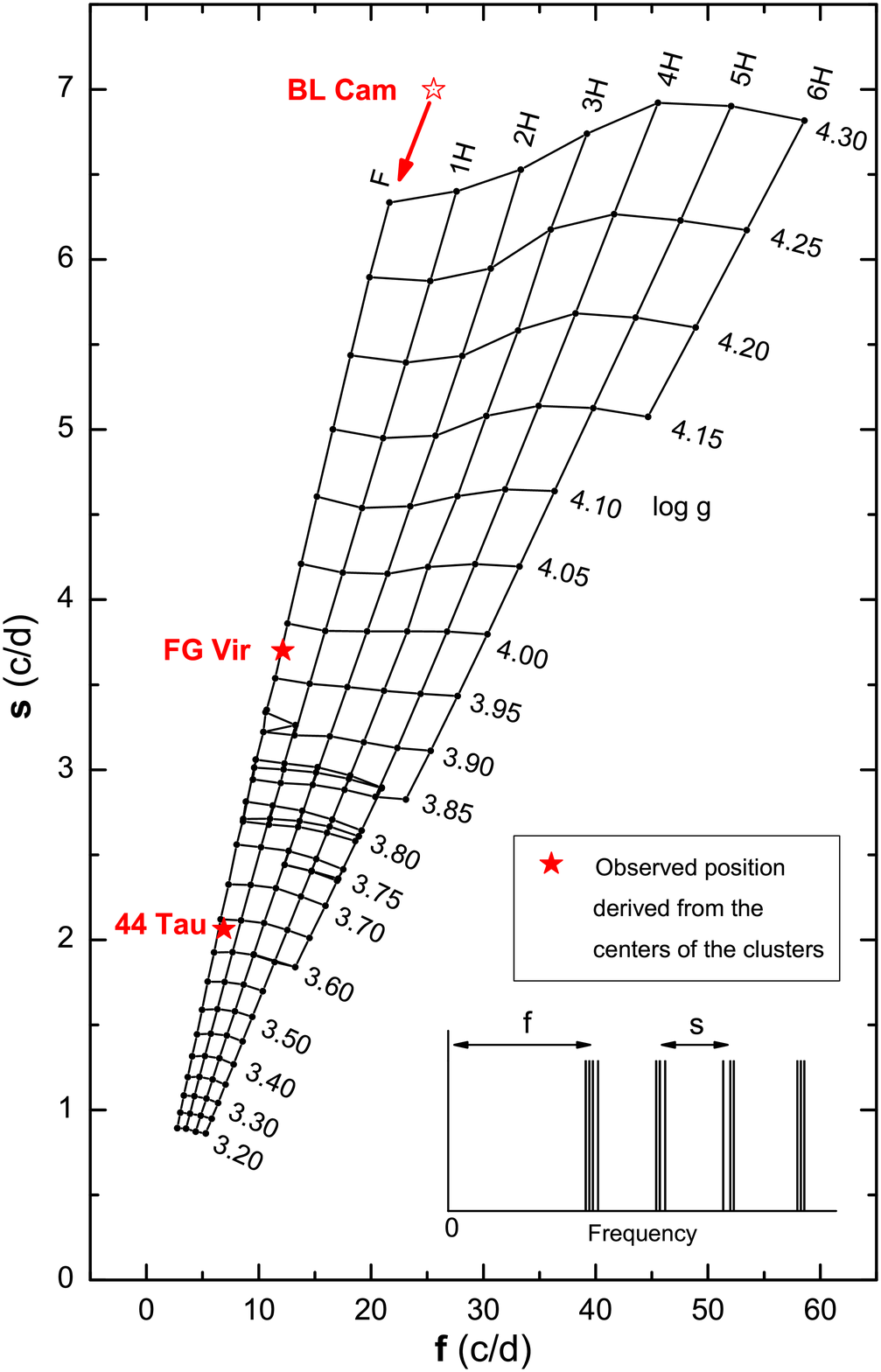}{Average separation of the radial modes, $s$, against the frequency of the lowest-frequency radial mode, $f$. We assume that each unstable radial mode represents the center of a cluster as illustrated on the bottom right. The grid makes it possible to determine $\log g$ and the order of the radial mode corresponding to the lowest frequency cluster. The asterisks correspond to observed values for the cluster centers of 44Tau, FG Vir and BL Cam.}{label}{!ht}{bb=-70 12 940 1162,width=115mm,clip}

Mode instability calculations show that the lowest-frequency cluster corresponds to the position of the radial fundamental mode only for the cool $\delta$ Scuti stars. For the hotter stars, instability shifts to higher radial orders. The position of blue instability borders for modes up to the sixth radial overtone in a HR diagram is given in Fig. 2. \emph{While the values of both the $s$ and $f$ parameters depend on the mean density and, therefore, the evolutionary stage of a star, the $f$ value also includes a temperature dependence.}

Fig. 3 shows the grid for the $\log g$ determination for stars in the $\delta$~Scuti domain. The grid points were derived from the position of unstable radial modes. The $s$ value was derived from the average spacing of all unstable radial modes, while $f$ is the frequency of the lowest unstable radial mode. The $s$ and $f$ values from different models with the same $\log g$ value were averaged to obtain the grid points.
As can be seen in Fig. 3, the transition between main-sequence models and post-main sequence models takes place at $\log g$ values between 3.90 and 3.75. 

We note that the s-f diagram  is in some respect similar to Petersen diagrams (period ratios of consecutive overtones
versus the longer period of each pair). Petersen diagrams also allow to determine the order of observed radial modes, as shown in Fig. 6 of Olech et al. (2005).

The grid presented in this paper was obtained with the standard values for chemical composition (X=0.70, Z=0.02, OPAL GN93) for nonrotating models. Since the frequencies and the instability of radial modes are affected by changes of metallicity, rotation, helium abundance and convection, we also computed corresponding models to test for all these effects. The detailed results of these tests will be presented in a forthcoming paper. We conclude that the accuracy of $\log g$ determination with the s-f diagram is comparable or even better than photometrical estimates.

In Fig. 3 the observed positions (derived from the centers of the clusters) of FG~Vir, 44~Tau and BL~Cam are shown. FG Vir and 44 Tau have normal chemical composition, whereas BL Cam is extremely metal-deficient (Z=0.0001). We computed pulsation models for BL Cam to determine the shift of the corresponding grid in $\log g$ accurately. The amount of the shift is shown in Fig. 3 by an arrow. 

If we are close to the asymptotic regime (oscillations in higher overtones), the spacings between frequency clusters can be two times smaller than the separation between consecutive radial  modes, because asymptotic $\ell=1$ modes lie halfway between radial modes. Consider a star with an observed frequency separation of 2 cd$^{-1}$ and a first frequency cluster at 10 cd$^{-1}$. If we misinterpret this frequency difference to correspond to the separation between $\ell$=0 and $\ell$=1 modes (as in the asymptotic case), the predicted radial separation would be 4 cd$^{-1}$. This value is located outside the grids shown in the s-f diagram, and no unstable modes are expected. Consequently, the incorrect value of 4 cd$^{-1}$ is ruled out so that 2 cd$^{-1}$ has to be the separation of radial frequencies. Any ambiguities may be ruled out this way and an incorrect $\log g$ determination is avoided.

\section{Conclusion}

In the observed pulsation spectra of well-studied $\delta$ Scuti stars a regular distribution of frequencies can be found. The detected nonradial frequencies tend to cluster in groups around radial modes. The comparison of the observations with theoretical pulsation models reveals that the cluster pattern may be due to trapping of modes in the stellar envelope. We present the s-f diagram. It relates the two parameters, viz., $f$, the frequency of the lowest radial mode,
and $s$, the mean spacing between the radial modes. Only linearly unstable modes are considered. For stars in the $\delta$~Scuti domain in the HRD the s-f diagram allows to infer the stellar $\log g$ value and to determine the order of the radial mode associated with the lowest-frequency cluster.

\acknowledgments{This investigation has been supported
by the Austrian Fonds zur F\"{o}rderung der wissenschaftlichen Forschung
and the Polish MNiI grant No.~1~P03D~021~28.}

\References{
\rfr  Breger, M., \& Lenz, P. 2008, A\&A, 488, 643
\rfr  Breger, M., Lenz, P., Antoci, V., et al. 2005, A\&A, 435, 955
\rfr  Bruntt, H., Su\'arez, J. C., Bedding, T. R., et al. 2007, MNRAS, 461, 619
\rfr  Dziembowski, W.~A., \& Kr\'{o}likowska, M. 1990, AcA, 40, 19
\rfr  Dupret, M. A., Grigahc\'ene, A., Garrido, R., Gabriel, M., Scuflaire, R. 2004, A\&A, 414, 17
\rfr  Lenz, P., Pamyatnykh A. A., Breger M., \& Antoci V. 2008, A\&A, 478, 855
\rfr  Olech, A., Dziembowski, W. A., Pamyatnykh, A. A., et al. 2005, MNRAS, 363, 40
\rfr  Rodr\'{i}guez, E., Fauvaud S., Farrell, J. A., et al. 2007, A\&A, 471, 255
}

\end{document}